\documentclass[twocolumn,aps,prl,superscriptaddress]{revtex4}
\usepackage[latin9]{inputenc}
\usepackage{amsmath}
\usepackage{amssymb}
\usepackage{graphicx}

\makeatletter
\@ifundefined{textcolor}{}
{%
 \definecolor{BLACK}{gray}{0}
 \definecolor{WHITE}{gray}{1}
 \definecolor{RED}{rgb}{1,0,0}
 \definecolor{GREEN}{rgb}{0,1,0}
 \definecolor{BLUE}{rgb}{0,0,1}
 \definecolor{CYAN}{cmyk}{1,0,0,0}
 \definecolor{MAGENTA}{cmyk}{0,1,0,0}
 \definecolor{YELLOW}{cmyk}{0,0,1,0}
}

\usepackage{epsfig}
\usepackage{xcolor}
\usepackage{tikz}

\makeatother
\begin{document}

\title{Localization driven superradiant instability}
\author{Honghao Yin}
\affiliation{Department of Physics, Capital Normal University,
Beijing 100048, China}
\author{Jie Hu}
\affiliation{Department of Physics, Capital Normal University,
Beijing 100048, China}
\author{An-Chun Ji}
\email{andrewjee@sina.com}
\affiliation{Department of Physics,
Capital Normal University, Beijing 100048, China}
\author{G. Juzeli\={u}nas}
\email{gediminas.juzeliunas@tfai.vu.lt}
\address{Institute of Theoretical Physics and Astronomy, Vilnius University,
 Saul\.etekio 3, LT-10257 Vilnius, Lithuania}
 \author{Xiong-jun Liu}
 \email{xiongjunliu@pku.edu.cn}
 \address{International Center for Quantum Materials, School of Physics, Peking University, Beijing 100871, China}
 \address{Collaborative Innovation Center of Quantum Matter, Beijing 100871, China}
\address{Beijing Academy of Quantum Information Science, Beijing 100193, China}
\address{Institute for Quantum Science and Engineering and Department of Physics, Southern University of Science and Technology, Shenzhen 518055, China}
\author{Qing Sun}
\email{sunqing@cnu.edu.cn}
\affiliation{Department of Physics, Capital
Normal University, Beijing 100048, China}
\date{\today }

\begin{abstract}
The prominent Dicke superradiant phase arises from coupling an ensemble of atoms to cavity optical field when external optical pumping exceeds a threshold strength. Here we report a prediction of the superrandiant instability driven by Anderson localization, realized with a hybrid system of Dicke and Aubry-Andr\'{e} (DAA) model for bosons trapped in a one-dimensional (1D) quasiperiodic optical lattice and coupled to a cavity.
Our central finding is that for bosons condensed in localized phase given by the DAA model, the resonant superradiant scattering is induced, for which the critical optical pumping of superradiant phase transition approaches zero, giving an instability driven by Anderson localization.
The superradiant phase for the DAA model with or without a mobility edge is investigated, showing that the localization driven superradiant instability is in sharp contrast to the superradiance as widely observed for Bose condensate in extended states, and should be insensitive to temperature of the system. This study unveils an insightful effect of localization on the Dicke superradiance, and is well accessible based on the current experiments.
\end{abstract}

\maketitle

Combining cold atomic gases with cavity quantum electrodynamics
\cite{Domokos2003JOSA, Asboth2005PRA, Nagy2006EPL, Mekhov2007NP,
Brennecke2007Nature, Colombe2007Nature, Nagy2008EPJD, Ritsch2013RMP,
Mekhov2012JPB} has provided a unique platform to explore exotic
quantum states in atom-cavity coupling systems \cite{Larson2008PRL,
Gopalakrishnan, Strack, Brennecke2008Science, Nagy2010PRL,
Gupta2007PRL, Keeling2010PRL, Kanamoto2010PRL,
Sun2011PRA,Padhi2013PRL, Landig2016Nature, Lode2017PRL}. In particular, the cavity field mediates an effective long-range
interaction between all atoms, and a prominent superradiant phase
with the atoms absorbing and emitting the photons collectively was
predicted in the notable Dicke model \cite{Dicke1954PR,Carmichael2007PRA}.
Such superradiance transition has been achieved
dynamically with a Bose-Einstein condensate (BEC) coupled to a
transversely pumped optical cavity
\cite{Baumann2010Nature,Baumann2011PRL, Mottl2012Science,
Brennecke2013PNAS, Leonard2017Nature}. Furthermore, for degenerate Fermi gases inside a cavity, the superradiance with enhancement by the Fermi surface nesting was predicted~\cite{Keeling2014PRL,Piazza2014PRL,Chen2014PRL}, and further the topological superradiant phases were also proposed~\cite{Pan2015PRL,Yu2018Front}. These studies
reveal the strong correlations between cavity photons and external
center-of-mass (COM) motion of atomic assembles in the dispersive
coupling regime \cite{Ritsch2013RMP}, with many exotic nonequilibrium
quantum behaviors having been uncovered in these open systems~\cite{Bakhtiari2015PRL, Klinder2015PNAS,Zheng2016PRL}.

The emergence of superradiance for BECs in a cavity
is typically associated with the formation of a self-organized supersolid
\cite{Baumann2010Nature,Baumann2011PRL}. The
disordered potential, if applied to the atoms, is expected to have significant effect on superradiance. In particular, the cavity-induced incommensurate lattice can induce the Bose-glass phases in Bose-Hubbard system as the optical pumping is strong enough \cite{Habibian2013PRL}, affect
localization transition of the atomic COM motion
\cite{Zhou2011PRA,Rojan2016PRA}, and lead to anomalous diffusion of the
atomic wavepackets \cite{Zheng2018PRA}. In these studies, the atoms are in ordered or extended states before the superradiance occurs. A question is, what happens if considering the coupling of an initially localized phase to cavity?

In this letter, we investigate a BEC in a localized phase given by a one-dimensional
quasi-periodical superlattice potential and coupled to a
transversely-pumped optical cavity, the latter providing an
effective long range interaction between the atoms. The
incommensurate quasi-periodical potential can lead to the Anderson
localization~\cite{Aubry1980,Roati2008Nature,Edwards2008PRL,Lahini2009PRL,
Fallani2007PRL,Modugno2009NJP,Modugno2010RPP,Lahini2010PRL,Biddle2009PRA,
Biddle2010PRL,Ganeshan2013PRL,Larcher2011PRA} and many-body localization which has attracted a considerable amount of interest recently~\cite{Schreiber2015Science,Ganeshan2015PRL,
Luschen2017arXiv, Li2017PRB}. 
In the extended regime, increasing the strength of the incommensurate potential can facilitate the tendency to the superradiant phase. Surprisingly, when the atoms enter the localized phase, we show that an unprecedented superradiant instability is driven by the resonant superradiant scatterings, for which the superradiance occurs at arbitrarily small optical pumping strength.

We consider a BEC inside a high-finesse optical cavity along the $x$-direction
(Fig.~1). The atoms experience a one-dimensional (1D) static bichromatic incommensurate potential
$V_{\rm static}(x)=V_1\cos^2(k_1x)+V_2\cos^2(k_2x+\phi)$ obtained by
superimposing two optical lattices, with $\phi$ a tunable relative phase, and are also illuminated by a standing-wave pumping laser
with the frequency $\omega_p$ in the $z$-direction.
The transverse confinement is sufficiently large so that the transverse motion
of atoms is suppressed. In the rotating frame, the Hamiltonian reads
$\hat{\mathcal{H}}=\int dx\hat{\psi}^{\dag}(x)\hat{H}\hat{\psi}(x)-\hbar\Delta_c
\hat{a}^{\dag}\hat{a}$, where $\Delta_c=\omega_p-\omega_c$ is the detuning
between the pump laser and the cavity field, $\hat{\psi}(x)$ and $\hat{a}$ are the
annihilation operators of the atom and cavity photon respectively. The full atomic Hamiltonian reads $\hat{H}=\hat{H}_0+V_{\rm dynamic}(x)$, with $\hat{H}_0=-\frac{\hbar^2}{2m}\nabla^2+V_{\rm static}(x)$. Here $V_{\rm dynamic}(x)=\hbar\eta(\hat{a}^\dag+\hat{a})\cos(k_cx)+\hbar U\hat{a}^\dag\hat{a}\cos^2(k_cx)$ is the cavity-assisted potential, $m$ is the atom mass and $\eta=\eta_0\cos(k_pz_0)$ describes the transverse pumping via atoms, with $\eta_0=g\Omega/\Delta_a$ being the strength of the interference between the pump lasers and the cavity field. Here also $U=g^2/\Delta_a$ is the atom-cavity
coupling strength, $\Delta_a=\omega_p-\omega_a$ denotes the detuning between the pumping laser and atomic transition frequency $\omega_a$, $\Omega$ is the strength (Rabi frequency) of the pumping laser, and $g$ is the single-photon Rabi frequency of the cavity mode.

\begin{figure}[t]
 \centering
 \includegraphics[width=0.4\textwidth]{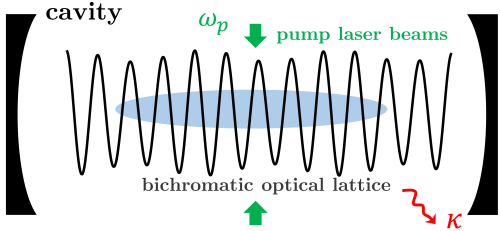}
 \caption{A schematic diagram of the system. A 1D
 BEC is placed inside a high-finesse optical cavity and driven by pump laser beams counter-propagating along the $z$-direction. The atoms are subjected to a 1D static bichromatic potential along the $x$-direction, and the transverse motion of atoms is suppressed by a large transverse confinement.}\label{f1}
\end{figure}

To better understand the model, we analyze first the situation without the pumping laser and
the cavity field. We choose
$V_1\cos^2(k_1x)$ ($V_1>V_2$) as the prime lattice, and the
secondary lattice $V_2\cos^2(k_2x)$ is relatively weaker.
In the tight-binding limit, the atomic Hamiltonian $\hat{\mathcal{H}}_0$ for the bichromatic
potential can be cast in the form of the  Aubry-Andr\'{e} (AA) model
\cite{Aubry1980}
\begin{equation}\label{e2}
\hat{\mathcal{H}}_{AA}=-J\sum_{j}\big(\hat{c}^{\dag}_j\hat{c}_{j+1}+h.c.\big)
  +\chi\sum_{j}\cos(2\pi \gamma
  j+\phi)\hat{c}^{\dag}_j\hat{c}_{j}.
  \end{equation}
Here $J$ is the tunneling matrix element between neighboring lattice sites and
the quasirandom disorder is induced by an additional incommensurate lattice,
characterized by the ratio of the lattice wave numbers $\gamma=k_2/k_1$ and
disorder strength $\chi$. For the maximally incommensurate
ratio $\gamma=(\sqrt{5}-1)/2$, the model undergoes an Anderson transition from
extended to localized states at $\chi/J=2$, beyond which all the states are localized. Such Anderson transition has been
well observed for non-interacting BECs \cite{Roati2008Nature,Edwards2008PRL}
and photonic crystals \cite{Lahini2009PRL}.
Beyond the tight-binding limit, corrections are
added to the AA model, leading to a general  Aubry-Andr\'{e} (GAA) model
Hamiltonian $\hat{\mathcal{H}}_{GAA}=\hat{\mathcal{H}}_{AA}+\hat{\mathcal{H}}^{\prime}$, with
\begin{eqnarray}\label{e3}
  \hat{\mathcal{H}}^{\prime}&=&J_2\sum_{j}\big(\hat{c}^{\dag}_j\hat{c}_{j+2}+h.c.\big)
  +\chi^{\prime}\sum_{j}\cos(4\pi \gamma j+2\phi)\hat{c}^{\dag}_j\hat{c}_{j}\nonumber\\
  &&+J^{\prime}\sum_{j}\cos\big[2\pi\gamma(j+\frac{1}{2})+\phi\big]\big(\hat{c}^{\dag}_j\hat{c}_{j+1}+h.c.\big),
\end{eqnarray}
where $J_2$ is the next-nearest-neighbor (NNN) hopping amplitude,
$J^{\prime}$ and $\chi^{\prime}$ are the correction parameters to the
tunneling parameter $J$ and disorder strength $\chi$, respectively. Unlike the AA model,
the GAA model may have an intermediate phase, where the localized and extended
eigenstates can coexist and separated by a single-particle mobility edge (SPME)~\cite{Li2017PRB}.

\begin{figure}
  \centering
  \includegraphics[width=0.5\textwidth]{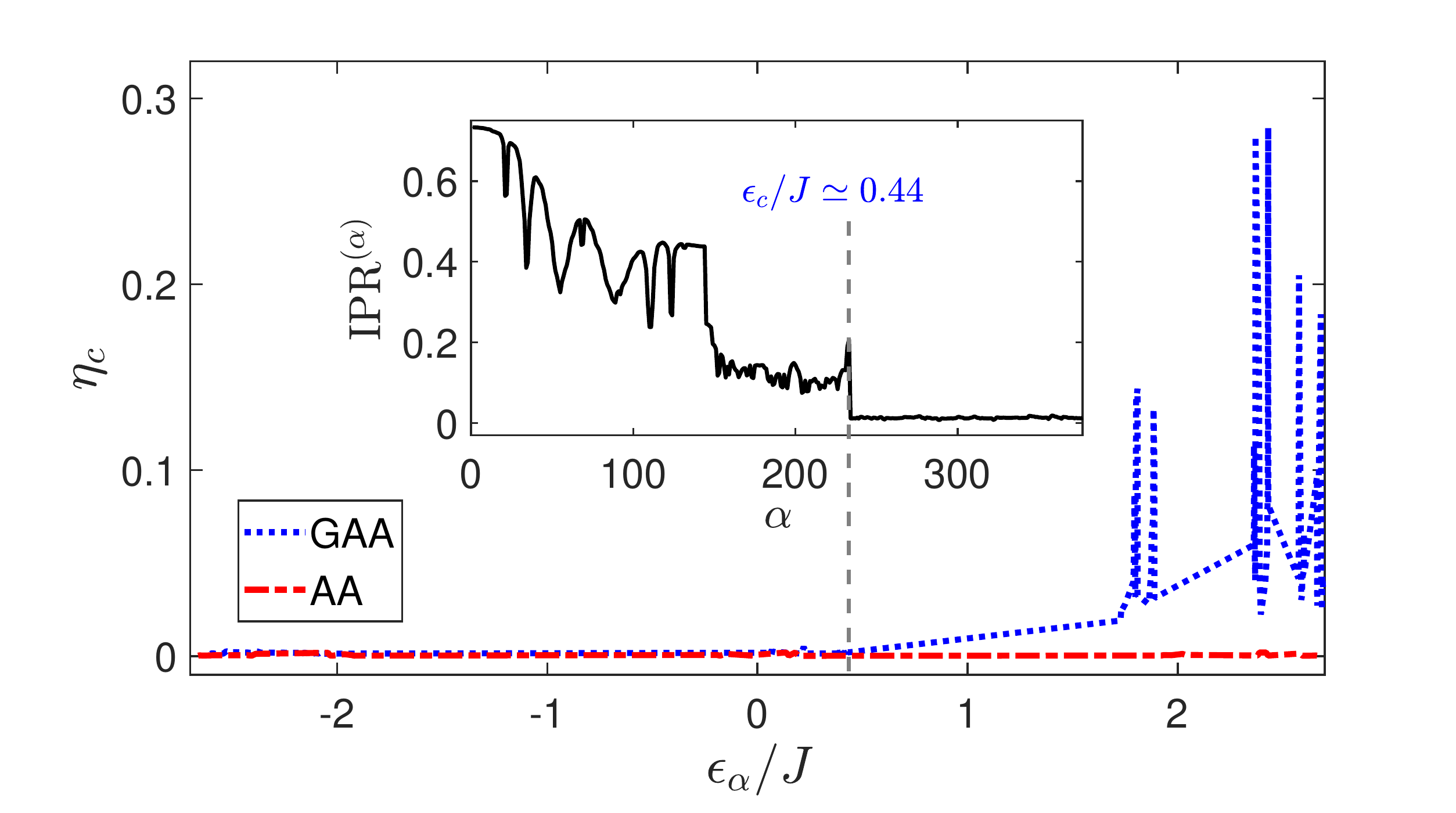}
  \caption{The critical pumping strength $\eta_c$ as a function of the energy $\epsilon_\alpha/J$ for atoms in different eigenstates at the disorder strength $\chi/J=2.1$ for GAA/AA model. The inset shows the inverse participation ratio (IPR) of the eigenstates in GAA model with a SPME ($\alpha\sim233$ marked by vertical dashed line). We have chosen $U/J=0.1$, $J^{\prime}/J=-0.23$, $J_2/J=0.072$, $\chi^{\prime}/J=-0.016$ for the GAA model (blue-dotted triangle) and the corresponding AA model (red-dash-dotted circle). Other parameters are $\gamma=\frac{233}{377}\simeq0.618$, $\gamma_c=0.8$, and $L=377$.}\label{f2}
\end{figure}

Our whole system is a hybrid Dicke and Aubry-Andr\'{e} (DAA) model, characterizing BEC in the quasi-periodic lattice and coupled to the cavity. The dynamics of the BEC and the cavity field can be captured by the master equation $\dot{\rho}=-i[\hat{\mathcal{H}},\rho]+\mathcal{L}\rho$ on the density matrix $\rho$, where $\mathcal{L}\rho=\kappa(2\hat{a}\rho\hat{a}^{\dag}-\hat{a}^{\dag}\hat{a}\rho-\rho\hat{a}^{\dag}\hat{a})$ is a Lindblad term to describe the cavity loss with a decay rate $\kappa$. Replacing the field operator by the c-number $a\equiv \langle\hat{a}\rangle$ yields $i\partial_t a=(-\Delta_c-i\kappa+Us_1)a+\eta s_0$. Here $s_0=\sum_j\cos(2\pi\gamma_c j)\langle\hat{c}^{\dag}_j\hat{c}_{j}\rangle$ and $s_1=\sum_j\cos^2(2\pi\gamma_c j)\langle\hat{c}^{\dag}_j\hat{c}_{j}\rangle$, with $\gamma_c=k_c/2k_1$ and $\langle\hat{c}^{\dag}_j\hat{c}_{j}\rangle$ the atomic density distribution. We seek for the steady-state solution by setting $\partial_t a=0$, and
\begin{equation}\label{pho}
  a=-\frac{\eta\sum_{j}\cos(2\pi\gamma_cj+\phi)\langle
  \hat{c}^{\dag}_j\hat{c}_j\rangle}{-\Delta_c-i\kappa
  +U\sum_{j}\cos^2(2\pi\gamma_cj+\phi)\langle
  \hat{c}^{\dag}_j\hat{c}_j\rangle}.
\end{equation}
Note that $\langle\hat{c}^{\dag}_j\hat{c}_{j}\rangle$ itself depends on the cavity-assisted potential, and should be determined self-consistently. In general, one can expect a transition from a ``normal" state with $a=0$ to a ``superradiant" state with $a\neq0$ by tuning e.g. the optical pumping strength $\eta$.

Figure~\ref{f2} shows the numerical results of the critical pumping strength $\eta_c$ versus the energy
of the BEC state for the GAA/AA model. For $\chi/J=2.1$, the GAA model gives an intermediate phase
with a SPME around the energy $\epsilon_c/J\simeq0.44$. As depicted in the inset of Fig.~\ref{f2}, the inverse participation ratio ${\rm IPR}^{(\alpha)}\equiv\sum_j|\phi^{\alpha }_j|^4/(\sum_j|\phi^{\alpha }_j|^2)^2$ vanishes
for extended states and becomes finite for localized states across the SPME. For the AA model, as
$\chi/J>2$, the system is in the localization phase with all the eigenstates being localized.

Our key observation is that an unprecedented localization driven superradiant instability of
the cavity field is obtained, i.e. the cavity field emerges spontaneously
for an arbitrarily small pumping strength. More exactly, whenever the BEC is in a localized state, no matter the ground state or an excited one, the superradiance can take place at a vanishing critical pumping strength. This is in sharp contrast to the extended state, where a finite pumping strength is
generally required~\cite{Baumann2010Nature,Baumann2011PRL,Mottl2012Science,Brennecke2013PNAS,Leonard2017Nature} (see also the blue-dotted line of the GAA model in Fig.~\ref{f2}). Since all the states in AA model with $\chi/J>2$ are localized and not thermalizable, the superradiant instability can be obtained for the BEC with any energy within the localized band, implying that this result is insensitive to the temperature.

To gain a deeper insight to the underlying physics, we analyze the superradiant
behaviour as the atomic wavefunction undergoes delocalization-to-localization transition.
We take the AA model for illustration. Diagonalizing the Hamiltonian (1), we have
$\hat{\mathcal{H}}_0=\sum_{\alpha}\varepsilon_{\alpha} \hat{c}^{\dag}_{\alpha}
\hat{c}_{\alpha}$, with $\varepsilon_{\alpha}$ and $\hat{c}_\alpha=\sum_{j}\hat{c}_{j}
\phi^j_{\alpha}$ being the eigenenergy and annilation operator of corresponding eigenstate $\phi_{\alpha}$. The total Hamiltonian of the system can then be rewritten as
\begin{eqnarray}\label{e3}
\hat{\mathcal{H}}&=&\sum_{\alpha}\varepsilon^{\prime}_{\alpha}
  \hat{c}^{\dag}_{\alpha}\hat{c}^{}_{\alpha}+\eta(\hat{a}^{\dag}
  +\hat{a})\sum_{\alpha\beta}(s^{}_{\alpha\beta}\hat{c}^
  {\dag}_{\alpha} \hat{c}^{}_{\beta}+h.c.)\nonumber\\
  &&+U\hat{a}^{\dag}\hat{a}\sum_{\alpha\beta}(h^{}_{\alpha\beta}
  \hat{c}^{\dag}_\alpha\hat{c}^{}_\beta+h.c.)-\Delta_{c}\hat{a}^{\dag}\hat{a},
\end{eqnarray}
which leads to a series of coupled motion equations
\begin{eqnarray}\label{e6}
  \!\!i\dot{a} &=&  -i\kappa a-\Delta_ca+Ua\sum_{\alpha\beta}(h^{}_{\alpha\beta}
   c^{\ast}_{\alpha} c^{}_{\beta}+h.c.)\nonumber\\
  &&+\eta\sum_{\alpha\beta}
  (s_{\alpha\beta} c^{\ast}_{\alpha} c^{}_{\beta}+h.c.)\,,\\
 \!\!i\dot{c}^{}_{\alpha}\!\! &=& \!\!\varepsilon^{\prime}_{\alpha}
  c^{}_{\alpha}\!+\!\eta(a^{\ast}+a)\sum_{\beta}s^{}_{\alpha\beta}c^{}_{\beta}\!+\!U|a|^2\sum_{\beta}h^{}_{\alpha\beta}c^{}_{\beta}
  \,.
\end{eqnarray}
Here $\epsilon^{\prime}_\alpha\equiv\epsilon_{\alpha}-\epsilon_0$ measures the eigenenergy from the lowest one ($\alpha=0$), 
$s_{\alpha\beta}\equiv\sum_j\phi^{\prime j\ast}_{\alpha} \cos(2\pi\gamma_c
j)\phi^{\prime j}_{\beta}$ and $h_{\alpha\beta}\equiv\sum_j \phi^{\prime j\ast}_{\alpha}
\cos^2(2\pi\gamma_cj)\phi^{\prime j}_{\beta}$ denote the scatterings between $\alpha$ and $\beta$ states. For the superradiant transition, we take that the atoms are condensed in the lowest state $\alpha=0$, and weakly scattered to other states when cavity field emerges. In this case only the scattering terms $s_{0\alpha}$ and $h_{0\alpha}$ are relevant. We define for convenience $s_{\alpha}\equiv
s_{0\alpha}= s^{*}_{\alpha 0}$ and $h_{\alpha}\equiv h_{\alpha0}=h^{*}_{\alpha 0}$. The case for atoms initially condensed in other states is similar.

\begin{figure}
  \centering
  \includegraphics[width=0.5\textwidth]{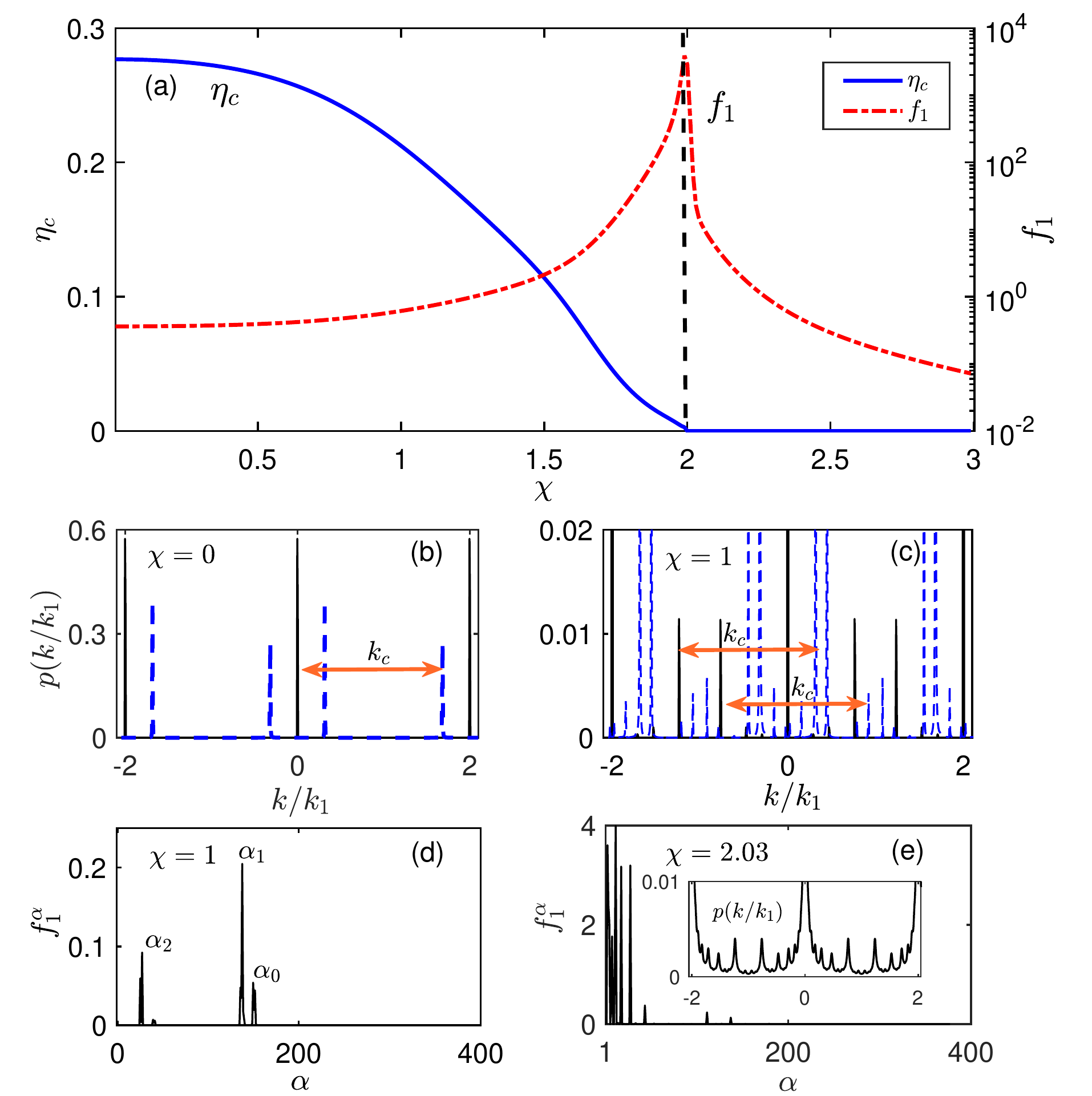}
  \caption{(a) The critical pumping field strength $\eta_c$ and the susceptibility
  $f_1$  as a function of the disorder strength $\chi$. Here, the parameters
  $L=377$, $N=100$, $\Delta_c/J=-1$, $\kappa/J=1$, and
  $\gamma_c=4/5$. The momentum distributions of the ground state
 (black solid) and and excited state (blue dashed) with disorder strength $\chi=0.0$ (b) and
 $1.0$ (c). Susceptibility $f^\alpha_1$ for the disorder strength $\chi/J=1$ (d) and $\chi=2.03$ (e). In the latter the term $f_1^0$ diverges and is not plotted. The inset shows the corresponding momentum distribution of the ground state.}\label{f4}
\end{figure}

Now we examine the superradiance phase transition. For the case
$\chi<2$, the wavefunctions of the states are extended, resembling the quasi-momentum states. One can find that the cavity field cannot induce self-scattering within the ground state and so $s_0=0$.
The critical value of the pumping strength reads
\begin{equation}\label{e8}
\eta_{c}=\sqrt{\frac{1}{4Nf_1}\frac{\kappa^2
+\bar{\Delta}^{\prime2}_c}{-\bar{\Delta}^{\prime}_c}},
\end{equation}
with $f_1=\sum_{\alpha}|s_\alpha|^2
/\varepsilon^{\prime}_\alpha$, $\bar{\Delta}^{\prime}_c=\Delta_c-2UNh^{}_{0}$, where $h_0=1/2$
gives a constant shift of cavity detuning. In the extended regime, the susceptibility $f_1$ is finite and the critical value of the pumping strength
determined by Eq.~({\ref{e8}}) is also finite. On the other hand, Fig.~3(a) shows that as the disorder
strength increases, the value $f_1$ increases rapidly (red curve), and the
superradiance tendency is strongly enhanced, with the critical
pumping field strength $\eta_c$ (blue curve) decreasing significantly with increasing the
disorder lattice potential. When increasing $\chi$ to the delocalization-to-localization transition point with $\chi=2$, the susceptibility $f_1$ diverges and the superradiance threshold becomes zero.

The unique role played by incommensurate lattice potential on the superradiance enhancement
becomes more transparent in the momentum space.
For the limit case with $\chi=0$ (no secondary disorder lattice),
the momentum distribution of the ground state,
$P(k/k_1)=|\sum_{j}\phi^j_0\exp(i\pi jk/k_1)|^2$, exhibits
primary peaks at $k=0$ and $2k_1$ (equivalent to $-2k_1$) in the first Brillouin zone
\cite{Edwards2008PRL,Lahini2009PRL,Fallani2007PRL,Modugno2009NJP,Modugno2010RPP}.
The momentum peaks of the $\alpha$-th excited
state are found to appear at $\pm\alpha k_1/L$ and
$\pm(2-\alpha/L)k_1$, which are shifted from the primary peaks of
the ground state by $\pm\alpha k_1/L$, with $\alpha$ being integers, as shown in
Fig.~\ref{f4}(b). Note that the pumping laser and
cavity field excite the atoms from the ground state to
the excited states. The scattering to the $\alpha$-th state contributes to the susceptibility
$f^\alpha_1\equiv|s_\alpha|^2 /\varepsilon^{\prime}_\alpha$. A
cavity photon carries a momentum $k_c$, so only the $\alpha_0$-th state with $\alpha_0=2L(1-\gamma_c)$ that matches
the cavity mode can be excited, giving a lattice version of Dicke model for noninteracting Bose gas.

When the secondary lattice is added, the momentum distributions of
the eigenstates are modified with the appearance of
new peaks. For the ground state, additional peaks
$\pm2(k_1-k_2)$, $\pm2k_2$ occur between the primary peaks. Accordingly, we show that  new momentum peaks of the excited states appear around the peaks of the
ground state by the distance $\alpha k_1/L$, see the dashed lines in
Fig.~\ref{f4}(c). In this case, besides the $\alpha_0$-th state, we
find two new excited states that take part in the atom-light
scattering process, with $\alpha_1=2L(\gamma_c-\gamma)$ and
$\alpha_2=2L(\gamma_c+2\gamma-2)$, see Fig.~\ref{f4}(d).
Compared to the $\alpha_1$-th state, which is a higher-excited
state near the $\alpha_0$-th state, the $\alpha_2$-th state is
located near the low-energy excitation regime, and can dominate the contribution to the
susceptibility. As the incommensurate lattice potential increases, more and more peaks arise in the momentum distributions of the ground and excited states, enhancing the contribution to the susceptibility and decreasing the critical pumping of the superradiant transition.

The nontrivial transition is obtained as the disorder strength $\chi$ approaches $2$, beyond which the eigenstates become localized in real space, but extended in the momentum space, namely, the momentum distribution of each state span the whole momentum space (see the inset of Fig.~\ref{f4}e for a reference). In such a situation, each localized state, including the ground
state with energy $\varepsilon_0$, can be scattered to itself by the cavity field via inducing transition between different momentum components  within the localized state. This gives rise to the resonant superradiant scattering. In consequence, the susceptibility
$f_1=\sum_\alpha f^\alpha_1$ diverges due to the contribution from the resonant self-scattering term $f^0_1$, for which superradiant instability is induced and the threshold pumping strength vanishes, as shown in Fig.~\ref{f4}(a).

We can find that $\eta_c\equiv 0$ in the whole
localized regime. The direct calculation shows that $s_{\alpha\alpha}$ is
finite for any localized state as $\chi>2$ and approaches $\cos(2\pi\gamma_cj_l)$ in the
deep localized regime, where $j_l$ is the central site of the localized
wavefunction. For a localized state distributing over a few
neighboring sites, the coherence of the light scattering by the atoms in such different sites can survive.
The backaction of the cavity makes the wavefunction of each state be more localized, giving rise to the self-organization at arbitrarily small pumping strength.

\begin{figure}
  \centering
  \includegraphics[width=0.5\textwidth]{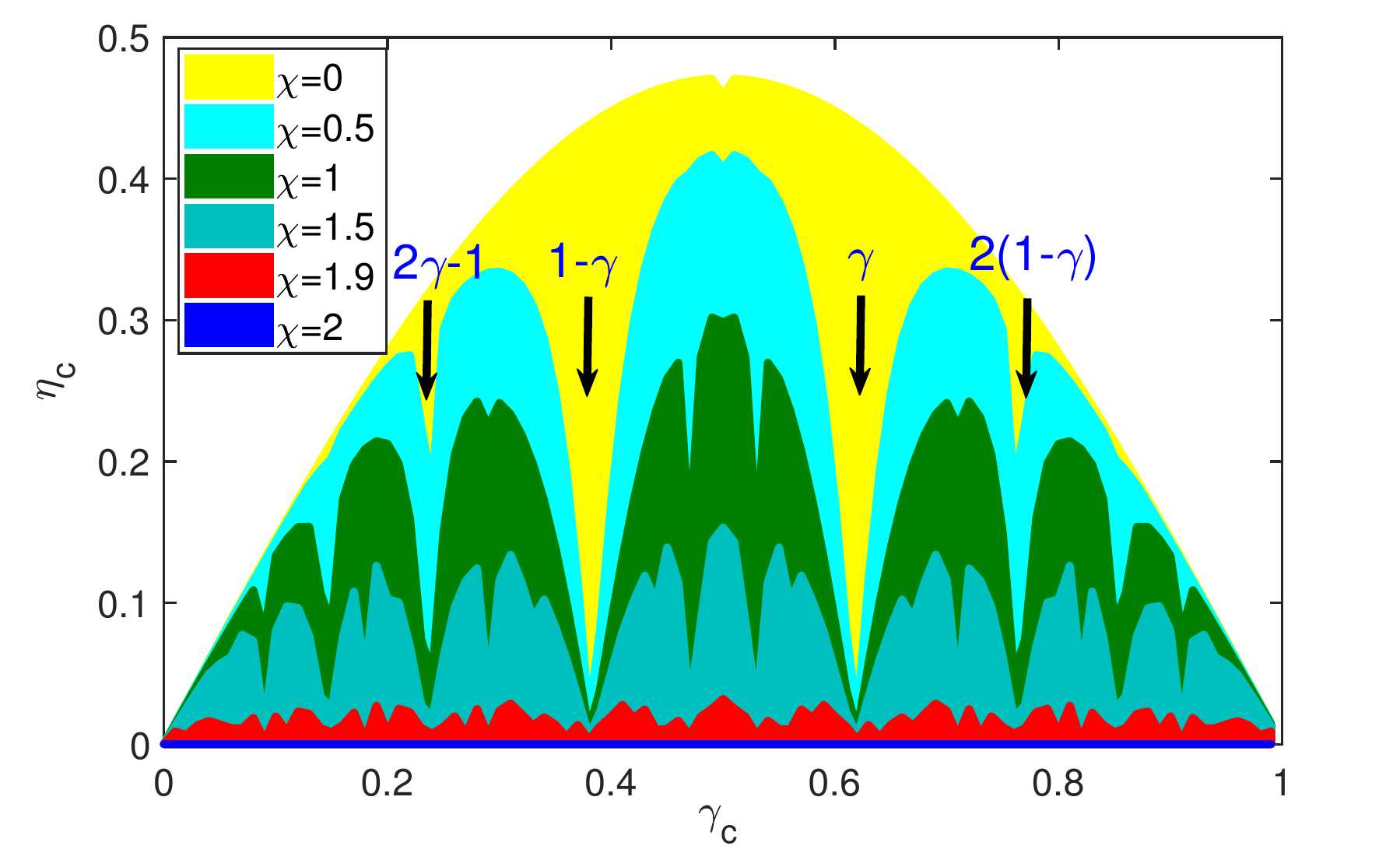}
  \caption{The diagram of the critical pumping field strength $\eta_c$  versus
$\gamma_c$ for different disorder strength $\chi$.
Here, the parameters are $L=377$, $N=100$, $\Delta_c/J=-1$, $\kappa/J=1$. }\label{f5}
\end{figure}

Finally we present in Fig.~\ref{f5} the dependence of critical pumping
field strength $\eta_c$ on the wave vector of cavity field, as characterized by $\gamma_c$, with different disordered potential strengths. When $\gamma_c$ matches with
the disorder potential, i.e. $\gamma_c\sim n\gamma$ or
$n(1-\gamma)\mod1$, with $n$ being a positive integer (see the arrows in Fig. 4), the critical pumping
strength drops more quickly with the disorder potential. This is because the cavity field enhances the disorder potential. Furthermore, by analyzing carefully the size effect from $L=300$ to $L=10000$ sites,
we find that the results are size-independent. Note that in real experiment, a weak harmonic
potential is needed to trap BEC, which however cannot affect the localization properties
substantially \cite{Edwards2008PRL}.
With the Gauss approximation of the
Wannier state, $J\sim (V_1/E_R)^{0.75}\exp(-\sqrt{V_1/E_R})$ and
$\chi\sim (V_2/E_R) \exp(-1/\sqrt{V_1/E_R})$ with the recoil energy
of the prime lattice $E_R=\hbar^2 k^2_1/(2m)$ \cite{Modugno2009NJP},
one can tune $\chi$ and $J$ by varying $V_2/V_1$.

In conclusion, we have predicted theoretically a novel superradiant instability by coupling the BEC in a localized phase to the cavity, in which the optical pumping threshold for the superradiance vanishes. The localization drives resonant superradiant scattering, in sharp contrast to the extended phases, for which the superradiance phase can occur at a vanishing pumping strength. The prediction is well accessible in the current experiments, and is expected to be valid in the many-body localization regime~\cite{Schreiber2015Science,Ganeshan2015PRL,
Luschen2017arXiv, Li2017PRB} which is achieved once interaction between atoms is included to the present DAA model. This works can open up an intriguing avenue in bridging the studies on the Dicke superradiance and the Anderson localization or many-body localization.

We acknowledge N. R. Cooper and H. Zhai for helpful discussions.
This work is supported by the National Natural Science Foundation
of China (11875195, 11404225, 11474205, 11504037, 11574008, 11761161003, 11825401), National Key R\&D Program of China (2016YFA0301604), and the Research
Council (Grant No. MIP- 086/2015). Q. Sun and A.-C. Ji also acknowledge the support by the foundation of Beijing Education Committees under Grants No. CIT\&TCD201804074 and No. KZ201810028043.

\end{document}